\documentclass[aps,pra,10pt,showpacs,twocolumn,groupedaddress,amsmath,amssymb]{revtex4-1}
\usepackage{bm}
\usepackage{graphicx}

\renewcommand{\paragraph}[1]{{\par\it #1.---}\ignorespaces}
\newcommand{\tr}{\ensuremath{\operatorname{tr}}}

\newcommand{\ev}[1]{\ensuremath{\langle #1 \rangle}}
\newcommand{\abs}[1]{\ensuremath{|#1|}}
\newcommand{\com}[2]{\ensuremath{[#1,\;#2]}}

\begin{document}
	\title{Quantum transport, master equations, and exchange fluctuations}
	\author{Robert Hussein}
	\author{Sigmund Kohler}
	\affiliation{Instituto de Ciencia de Materiales de Madrid, CSIC, Cantoblanco, E-28049 Madrid, Spain}
	\date{\today}
	
	\begin{abstract}
		
		We investigate to which extent a many-body Bloch-Redfield master equation
		description of quantum transport is consistent with the exact generalized
		equilibrium conditions known as exchange fluctuation theorems.  Thereby we
		identify a class of master equations for which this is the case.  Beyond
		this class, we find deviations which exhibit characteristic scaling laws as
		functions of the dot-lead tunneling, the inter-dot tunneling, and the
		temperature.  These deviations are accompanied by an increase of lead
		energy fluctuations inherent in the Bloch-Redfield equation beyond
		rotating-wave approximation.  We illustrate our results with numerical data
		for a double quantum dot attached to four leads.
		
	\end{abstract}
	
	\maketitle

	Exchange fluctuation theorems are exact relations between probabilities for
	non-equilibrium transitions that start from a Gibbs state and reflect the
	time reversibility of the microscopic equations of motion
	\cite{Esposito2009a, Campisi2011a}.  Frequently they are expressed by the
	statistics of work performed at a system upon time-dependent parameter
	variation.  A variant thereof concerns charge and heat exchange in quantum
	transport between leads \cite{TobiskaPRB2005a,AndrieuxJSM2006a,Saito2007b, Saito2008a} and can be verified
	experimentally \cite{NakamuraPRL2010a,UtsumiPRB2010a,SairaPRL2012a}.  Taylor expansion of
	these exchange fluctuation theorems at equilibrium provides relations
	between transport coefficients such as the Johnson-Nyquist relation
	\cite{TobiskaPRB2005a,AndrieuxJSM2006a,Saito2008a}.
	
	Theoretical studies of quantum transport often rely on approximations such
	as perturbation theory in the tunneling between system and electron
	reservoirs to obtain a master equation approach
	\cite{Gurvitz1996a}.  It has been demonstrated that a careless application
	of master equations may predict spurious currents at equilibrium
	\cite{Novotny2002a} and thus may violate fluctuation theorems.  The
	validity of exchange fluctuation theorems has been verified for master
	equation descriptions of various specific situations \cite{Sanchez2010a,
	BulnesCuetara2011a, Golubev2011a, BulnesCuetara2013a, Utsumi2010a,
	Lopez2012a}.  Still the question arises whether any general statement
	for a whole class of master equations is possible.
	
	A widely employed Markovian master equation for quantum systems weakly coupled
	to environmental degrees of freedom is provided by the Bloch-Redfield formalism
	\cite{Redfield1957a}.  Being originally derived for dissipative quantum
	mechanics, it can be generalized straightforwardly to quantum transport, e.g.,
	to coupled quantum dots in contact with electron reservoirs.  Moreover, it is
	equivalent to various common master equations.
	In this work, we demonstrate that the Bloch-Redfield master equation is
	consistent with exchange fluctuation theorems only to some extent (it does
	not predict spurious equilibrium currents and maintains the Johnson-Nyquist
	relation), while it fully agrees only after a rotating-wave approximation
	(RWA).  Some previous results \cite{Sanchez2010a, BulnesCuetara2011a,
	Golubev2011a, BulnesCuetara2013a} emerge as limiting cases of our generic
	statements.
	Moreover, we predict for the fluctuation theorem violation of the
	Bloch-Redfield equation a scaling behavior which we confirm by a numerical
	study.
	\begin{figure}[b]
		\includegraphics{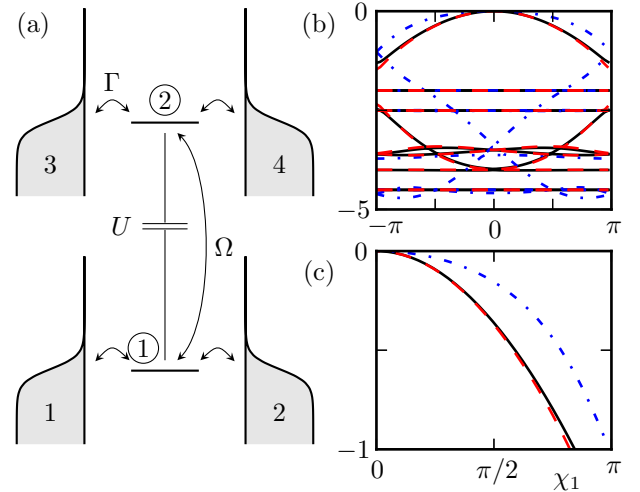}\caption{(a) Double quantum dot in contact with four leads
		$\alpha=1,\ldots,4$, used to exemplify our analytical results and the
		scaling of the deviations from the exact exchange fluctuation theorem
		\eqref{FT}.
		(b) Real part of the eigenvalues of the Liouvillians $\mathcal
		L_{\bm\chi,0}$ (solid lines), $\mathcal L_{-\bm\chi-i\beta\bm\mu,i\beta}$
		(dashed, hidden by solid lines), and $\mathcal L_{-\bm\chi-i\beta\bm\mu,0}$
		(dash-dotted) [see Eq.~\eqref{L}] as function of $\chi_1$ while all other
		$\chi_\alpha=0$ for inter-dot tunneling $\Omega=0.75\Gamma$, temperature
		$k_BT=0.1\Gamma$, onsite energies $\epsilon_1=2\,\epsilon_2=\Gamma$, and
		chemical potentials $\mu_1=-\mu_2=-\mu_3=-\mu_4=0.25\,\Gamma$.
		(c) Enlargement of panel b revealing the slight difference between
		$\mathcal L_{\bm\chi,0}$ and $\mathcal L_{-\bm\chi-i\beta\bm\mu,i\beta}$.
		\label{fig:setup}
		}
	\end{figure}
	
	\paragraph{Dot-lead model and exchange fluctuation theorem}
	We consider a transport setup of the type sketched in
	Fig.~\ref{fig:setup}(a) and
	modeled by the Hamiltonian $H = H_S + V + \sum_\alpha H_\alpha$, where $H_S$
	describes a central system, henceforth referred to as ``quantum dots''.
	Notably, in contrast to Refs.~\cite{Esposito2007a, Esposito2009a}, our
	system Hamiltonian $H_S$ may contain Coulomb repulsion terms which in most
	quantum dots represent the largest energy scale.  Thus for the
	decomposition of the density operator, we will have to work in a
	many-body basis.
	
	The other constituents of our system are leads modeled as free electrons
	with the Hamiltonian $H_\alpha = \sum_q \epsilon_{\alpha q} c_{\alpha q}^\dagger
	c_{\alpha q}$, where $c_{\alpha q}^\dagger$ creates an electron in mode $q$ of
	lead $\alpha$ with energy $\epsilon_{\alpha q}$.  Initially the leads are in a Gibbs
	ensemble at a common temperature $T$, while the chemical potentials
	$\mu_\alpha$ are shifted from their equilibrium values $\mu_\alpha=0$ by
	externally applied voltages.  This implies the expectation
	values $\langle c_{\alpha q}^\dagger c_{\alpha' q'}\rangle =
	\delta_{\alpha\alpha'} \delta_{qq'} f(\epsilon_{\alpha q}-\mu_\alpha)$ with the Fermi
	function $f(x) = [\exp(\beta x)+1]^{-1}$ and the inverse temperature
	$\beta = 1/k_BT$.
	Each lead $\alpha$ is tunnel coupled to one quantum dot $n_\alpha$ via a
	Hamiltonian $V_\alpha = \sum_q V_{\alpha q} c_{\alpha q}^\dagger
	c_{n_\alpha} + \text{H.c.}$, which is fully determined by the spectral
	density $\Gamma_\alpha(\epsilon) = 2\pi\sum_q |V_{\alpha q}|^2
	\delta(\epsilon-\epsilon_{\alpha q})$.  In our numerical calculations, we
	assume within a wide-band limit energy-independent couplings,
	$\Gamma_\alpha(\epsilon) \equiv \Gamma_\alpha$, while our analytical results are
	valid beyond.
	
	For the computation of the stationary current and its low-frequency
	fluctuations, we employ the cumulant generating function $Z(\bm\chi) =
	\lim_{t\to\infty}\frac{\partial}{\partial t}\ln\langle e^{i\bm\chi\cdot\bm
	N}\rangle$ \cite{Bagrets2003a}, which implicitly depends on the chemical
	potentials $\mu_\alpha$ [the vector components refer to the different
	leads, $(\bm x)_\alpha \equiv x_\alpha$].  Its idea is to generate the
	lead electron number operator $N_\alpha$ via a derivative with
	respect to the counting variable $\chi_\alpha$, while the time derivative
	turns number cumulants into current cumulants.  Taylor expansion of
	the generating function at $\bm\chi = \bm\mu = \bm0$ yields (particle)
	transport coefficients, i.e., derivatives of the current and its cumulants
	with respect to the applied voltages.  In particular, $I_\alpha = (\partial
	Z/\partial i\chi_\alpha)|_{\bm\chi=\bm\mu=\bm0}$, the conductance
	$G_{\alpha,\beta} = -(\partial^2 Z/\partial i\chi_\alpha
	\partial\mu_\beta)|_{\bm\chi=\bm\mu=\bm0}$, while the zero-frequency limit
	of the current correlation function $\langle
	I_\alpha,I_\beta\rangle_{\omega\to0}$ reads $S_{\alpha\beta} =
	(\partial^2/\partial i\chi_\alpha\partial i\chi_\beta) Z|_{\bm\chi=\bm0}$
	\cite{Blanter2000a}.
	
	Using an exact formal solution of the dot-lead dynamics, one can
	demonstrate that the cumulant generating function obeys the exchange
	fluctuation theorem \cite{Saito2008a}
	\begin{equation}
		\label{FT}
		Z(\bm\chi) = Z(-\bm\chi-i\beta\bm\mu) ,
	\end{equation}
	Its practical use is to derive relations between different
	transport coefficients.  To first order, $I_\alpha=0$, while to second
	order one, e.g., obtains the Johnson-Nyquist relation
	$2k_BT G_{\alpha,\alpha} = S_{\alpha\alpha}$.  For a proof of
	Eq.~\eqref{FT} \cite{Saito2008a}, one introduces a counting variable $\xi$
	for the total lead \cite{Saito2007a, Ren2010a, Nicolin2011a} energy to
	obtain the relation $Z(\bm\chi,\xi) =
	Z(-\bm\chi-i\beta\bm\mu,-\xi+i\beta)$.  Then one argues that, provided that
	the energy of the central system is negligible, the total lead energy is
	conserved and, thus, $Z$ is independent of $\xi$.
	In the following, we explore up to which extent a Bloch-Redfield theory for
	quantum transport complies with this exact statement.
	
	\paragraph{Bloch-Redfield master equation}
	Within second-order perturbation theory for the dot-lead tunnel coupling
	$V$, we obtain for the reduced system density operator $\rho$ the Markovian
	master equation \cite{Redfield1957a} (for ease of notation, we set
	$\hbar=1=e_0$ and consider particle currents)
	\begin{equation}
		\dot\rho = -i[H_S,\rho] -\frac{1}{2} \int\nolimits_{-\infty}^{+\infty} d\tau
		\tr_\text{leads}[V,[\tilde V(-\tau),\rho\otimes\rho^\text{leads}_{\bm\mu}]] ,
		\label{ME}
	\end{equation}
	where $\rho_{\bm\mu}^\text{leads} \propto \exp[-\beta\sum_\alpha(H_\alpha
	-\mu_\alpha N_\alpha)]$ and $\tilde V$ is the interaction picture version
	of the tunnel Hamiltonian with respect to $H_S+\sum_\alpha H_\alpha$.  In order
	to achieve this form, we have symmetrized the time integral.  This corresponds to
	neglecting principal parts, which can be justified by renormalization arguments.
	
	In order to obtain the generating function within the Bloch-Redfield
	approach, $Z_\text{BR}$, we multiply in Eq.~\eqref{ME}
	the density operator by $\exp(i\bm\chi\cdot\bm N+i\xi\sum_\alpha
	H_\alpha)$, which yields
	$\dot\rho_{\bm\chi,\xi} = \mathcal L_{\bm\chi,\xi} \rho_{\bm\chi,\xi}$ with
	the generalized Liouvillian
	\begin{equation}
		\label{L}
		\mathcal L_{\bm\chi,\xi}
		= -i[H_S,\rho]-\mathcal D
		+\sum_\alpha\big[e^{-i\chi_\alpha}\mathcal J_\alpha^\text{in}(\xi)
		+e^{i\chi_\alpha}\mathcal J_\alpha^\text{out}(\xi) \big],
	\end{equation}
	where $\mathcal J_\alpha^\text{in/out}$ describe dot-lead tunneling, while
	$\mathcal{D}$ subsumes all other dissipative terms.  For vanishing counting
	variables, $\mathcal L_{\bm 0,0}=\mathcal{L}$, the physical Liouvillian.
	Since $\tr\rho_{\bm\chi,\xi}$ is the moment generating function for the
	leads electron number, the current cumulant generating function reads
	$Z_\text{BR}(\bm\chi,\xi) = \frac{\partial}{\partial t}
	\ln\tr\rho_{\bm\chi,\xi}$.  In the long time limit, the r.h.s.\ of this
	expression becomes identical to the eigenvalue of $\mathcal
	L_{\bm\chi,\xi}$ with smallest real part, which reduces the computation of
	current cumulants to an eigenvalue problem \cite{Bagrets2003a}.
	
	We cope with the interaction picture operator in the master
	equation \eqref{ME} by decomposing the density operator into the many-body
	eigenstates of the quantum dots, $\{|a\rangle\}$, where $H_S|a\rangle =
	E_a|a\rangle$.  Since the counting variables appear only in combination
	with the jump terms $\mathcal{J}_\alpha^\text{in/out}$, we restrict the
	discussion to these terms.  Their eigenbasis representation reads
	\begin{equation}
		\label{J}
		\begin{split}
			[\mathcal J_\alpha^\text{in}(\xi)]_{ab, a'b'}
			= {}&
			\frac{1}{2}\langle a|c_{n_\alpha}^\dagger|a'\rangle
			\langle b'|c_{n_\alpha}|b\rangle
			\\ & \times
			\big\{ F_\alpha^{<}(E_a-E_{a'}) e^{-i(E_a-E_{a'})\xi}
			\\ & \quad
			+F_\alpha^{<}(E_b-E_{b'}) e^{-i(E_b-E_{b'})\xi}
			\big\}
		\end{split}
	\end{equation}
	with the lesser and greater lead correlation function $F_\alpha^{<}(t) = \sum_q
	\abs{V_{\alpha q}}^2 \langle c_{\alpha q}^\dagger(0) c_{\alpha q}(t)\rangle =
	F_\alpha^{>}(t-i\beta) e^{\beta\mu_\alpha}$ \cite{Kubo1957a, Martin1959a}.
	Its Fourier representation in the wide-band limit reads
	$F_\alpha^{<}(\epsilon) = \Gamma_\alpha f(\epsilon-\mu_\alpha) =
	\Gamma_\alpha - F_\alpha^{>}(\epsilon)$, while the corresponding
	tunneling-out operators $[\mathcal J_\alpha^\text{out}(\xi)]_{ab, a'b'}$
	follow from the replacement $\{c_n,F^{<}(\epsilon)\} \to
	\{c_n^\dagger,F^{>}(-\epsilon)\}$.  Notice the dependence on energy
	differences of the many-body states, $E_a-E_{a'}$.  Only for non-interacting
	systems, this difference becomes a single-particle energy.
	
	A first glance of the results derived below, is provided by the spectra of
	$\mathcal L_{\bm\chi,\xi}$, $\mathcal L_{-\bm\chi-i\beta\bm\mu,-\xi+i\beta}$,
	and $\mathcal L_{-\bm\chi-i\beta\bm\mu,\xi}$ at $\xi=0$
	[Figs.~\ref{fig:setup}(b) and \ref{fig:setup}(c)].
	One notices that the former and the latter clearly disagree, which
	demonstrates that for $Z_\text{BR}$, it is not sufficient to consider only
	the number counting variable $\bm\chi$.  Thus, $Z_\text{BR}$ does not
	fulfill Eq.~\eqref{FT}, i.e., the full Bloch-Redfield equation violates the
	exchange fluctuation theorem.  When also the energy counting variable is
	substituted as $\xi\to -\xi+i\beta$, the difference between the spectra
	becomes significantly smaller, which indicates that the fluctuation
	theorem violation relates to the total lead energy.
	
	\paragraph{RWA master equation for many-body states}
	If after an irrelevant transient stage, the density operator becomes
	practically diagonal in the energy basis, one may employ the RWA ansatz
	$\rho_{ab} = P_a\delta_{ab}$, where the populations $P_a$ obey $\dot P_a =
	\sum_{a'} w_{a\leftarrow a'} P_{a'}$.  The transition rates $w_{a\leftarrow
	a'}$ consist of the tunnel-in contributions for each lead,
	\begin{align}
\nonumber
w^{\alpha,\text{in}}_{a\leftarrow a'} & (\bm\chi,\xi)
= [\mathcal{J}_\alpha^\text{in}]_{aa,a'a'} \\
={}& |\langle a|c_{n_\alpha}^\dagger|a'\rangle|^2
  e^{i\chi_\alpha} e^{-i(E_a-E_{a'})\xi}
  F_\alpha^{<}(E_a-E_{a'}) ,
\label{w}
	\end{align}
	and the corresponding $w_{a\leftarrow a'}^{\alpha,\text{out}}$.
	Using $F_\alpha^{<}(\epsilon) e^{\beta(\epsilon-\mu_\alpha)} =
	F_\alpha^{>}(\epsilon)$ \cite{Kubo1957a, Martin1959a}, we find
	$w_{a'\leftarrow a}(\bm\chi,\xi) = w_{a\leftarrow a'}(-\bm\chi-i\beta\bm\mu,-\xi+i\beta)$, i.e., the substitution $\bm\chi,\xi
	\rightarrow -\bm\chi-i\beta\bm\mu,-\xi+i\beta$ corresponds to the transposition
	of the RWA Liouvillian.
	Moreover, the $\xi$-dependence can be removed via the similarity
	transformation $w \rightarrow S^{-1} w S$ with $S_{a,a'} = \delta_{aa'}
	e^{iE_a\xi}$.  Since both transposition and the transformation with $S$
	leave the spectrum unchanged, we can draw two conclusions for
	the generating function being the lowest eigenvalue: First,
	$Z_\text{RWA}$ is $\xi$-independent which implies that the lead energy is
	conserved in the long-time limit.  Second, $Z_\text{RWA}(\bm\chi)$ fulfills
	Eq.~\eqref{FT}.
	
	\paragraph{RWA class of master equations}
	The above statement about the Bloch-Redfield master equation in RWA can be
	applied to master equations that are seemingly not of that form.  Moreover,
	the cases of vanishing Coulomb interaction and of infinitely strong
	repulsion emerge as single-particle limits of our statements.  In that
	sense, we can identify a whole ``RWA class'' of master equations for which
	Eq.~\eqref{FT} holds.
	
	A most relevant case is a master equation for capacitively coupled, but
	electrically isolated quantum dots, each modeled as single level.  Owing to
	the lack of coherent tunneling, the Hamiltonian of this system is diagonal
	in the onsite basis, while no quantum coherence emerges.  Thus,
	off-diagonal density matrix elements vanish exactly, so that the resulting
	master equation in a localized basis assumes the form of the RWA limit of
	the Bloch-Redfield equation.  Recently, the validity of the exchange
	fluctuation theorem has been exemplified for various particular situations
	of this kind \cite{Sanchez2010a, BulnesCuetara2011a, Golubev2011a,
	Lopez2012a, BulnesCuetara2013a}.  They represent special cases of our
	generic statement.
	
	Moreover, there are limits in which our many-body master equation becomes
	in fact a single-particle particle equation.  This is naturally the case for very
	strong inter-dot Coulomb repulsion, such that at most one electron can
	enter the system.  Then only eigenstates with one electron play a role and
	the energy differences in the jump operator \eqref{J} become single
	particle energies.  In the opposite limit of non-interacting electrons, the
	many-body states $|a\rangle$ are Slater determinants of single-particle
	states, while all $E_a$ are sums of single-particle energies, a case that
	has been considered, e.g., in Ref.~\cite{Esposito2009a}.  Again only the
	single-particle energies appear in the decomposition of $\mathcal{J}$.  We
	emphasize that genuine many-body effects or correlation effects typically
	emerge in between these regimes and, thus are beyond these limits.
	
	\paragraph{Exchange fluctuation theorem violation}
	\begin{figure*}[tb]
		\includegraphics{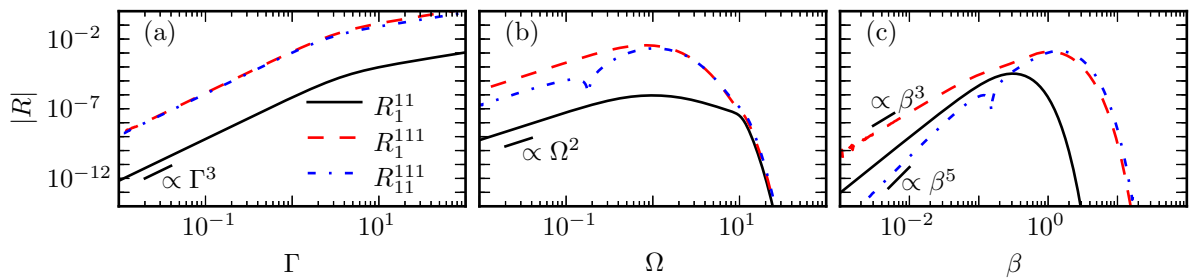}\caption{Violation of the exchange fluctuation theorem by the Redfield
		master equation beyond RWA for the quadruple quantum dot sketched in
		Fig.~\ref{fig:setup}(a) as function of (a) the dot-lead coupling $\Gamma$,
		(b) the inter-dot tunneling $\Omega$, and (c) the inverse temperature
		$\beta=1/k_BT$ for the default parameters $\Gamma=0.5\,\Omega=k_BT$ and
		$\epsilon_\alpha=\mu_\alpha=0$.  The scaling behavior verifies the
		conjecture \eqref{scaling} for the selected generalized Casimir-Onsager relations
		$R_{1}^{11}=0$ (solid line), $R_{1}^{111}=0$ (dashed), and $R_{11}^{111}=0$
		(dash-dotted).
		}
		\label{fig:R}
	\end{figure*}%
	Having seen that the full Bloch-Redfield equation violates the fluctuation
	theorem, we turn to a quantitative analysis of the deviations.  To this
	end, we introduce as measure the $(m+n)$th order Taylor coefficients of the
	difference between the terms appearing in Eq.~\eqref{FT},
	\begin{equation}
		\label{R}
		R^{\alpha_1\cdots\alpha_m}_{\beta_1\cdots\beta_n}
		=
		\frac{(-i)^m\partial^{m+n}}{\partial\chi_{\alpha_1} \cdots
		\partial\mu_{\beta_n}}
		\big\{ Z_\text{BR}(\bm\chi) - Z_\text{BR}(-\bm\chi-i\beta\bm\mu)\big\}
		\Big|_{\bm\chi=\bm\mu=\bm0} ,
	\end{equation}
	which are constructed such that they vanish if the exchange fluctuation
	theorem is fulfilled.
	Notice that $Z_\textrm{BR}$ possesses also an implicit $\bm\mu$-dependence,
	so that generally the contribution of the first term does not vanish.
	Since, the r.h.s\ of Eq.~\eqref{R} consists of derivatives of current
	cumulants evaluated at equilibrium $\bm\mu=\bm0$, the fact that
	$R_{\alpha_1\cdots\alpha_m}^{\beta_1\cdots\beta_n}$ must vanish provides a
	relation between transport coefficients \cite{Saito2008a}.  For example,
	the mentioned Johnson-Nyquist relation is of second order and reads
	$R_\alpha^\alpha = 0 = \beta S_{\alpha\alpha}-2G_{\alpha,\alpha}$.
	This rather important relation represents an interesting special case of
	Eq.~\eqref{R} because it is fulfilled also by the full Bloch-Redfield
	equation beyond RWA, as we prove in Appendix~B.
	
	Before entering numerical calculations, we like to conjecture the scaling
	behavior of the deviations \eqref{R} as function of (i) the incoherent
	tunnel rates $\Gamma$ and (ii) the coherent tunnel coupling $\Omega$.  In each
	case, we depart from a limit in which the fluctuation theorem~\eqref{FT} is
	fulfilled, so that all $R$ indeed vanish.  Concerning (i) we recall that
	the master equation \eqref{ME} is based on a perturbation theory in the
	dot-lead coupling which cannot capture the Lorentzian broadening of the
	quantum dot resonance denominator $\propto(\epsilon^2+\Gamma^2)^{-1}$.
	Thus, corrections to the exact equilibrium density matrix should be of the
	order $\Gamma^2$.  Moreover, since all transport coefficients inherit a
	prefactor $\Gamma$ from the jump operators [see Eq.~\eqref{J}], we expect
	$R\propto \Gamma^3$.  For case (ii) we notice that for $\Omega=0$, no
	coherent tunneling is present and the full Bloch-Redfield falls into the
	RWA class identified above, so that the fluctuation theorem holds exactly.
	Since expectation values typically depend only on even powers of tunnel
	matrix element $\Omega$, we anticipate deviations of order
	$\Omega^2$.
	Assuming that the deviations from $R=0$ depend on the smaller of both
	parameters, we can conjecture the generic behavior
	\begin{equation}
		\label{scaling}
		R \propto
		\begin{cases}
			\Gamma^3 & \text{for $\Gamma\ll\Omega$,} \\
			\Omega^2 & \text{for $\Omega\ll\Gamma$.}
		\end{cases}
	\end{equation}
	
	For the verification of this hypothesis for systems such as the one
	sketched in Fig.~\ref{fig:setup}, we have to derive a scheme for the
	computation of transport coefficients to high orders.  For this purpose, we
	generalize an iteration scheme for the computation of current cumulants
	\cite{Flindt2008a} to the computation of derivatives with respect to the
	chemical potentials $\mu_\alpha$ and the presence of an energy counting
	variable $\xi$.  The method is based on the fact that the transport
	coefficients are Taylor coefficients of the generating function in the
	variables $\chi_\alpha$, $\xi$, and $\mu_\alpha$, which can be computed
	iteratively by Rayleigh-Schr\"odinger perturbation theory, see Appendix~A.
	
	Figures \ref{fig:R}(a) and \ref{fig:R}(b) depict the scaling behavior of
	three different deviations as functions of $\Gamma$ and $\Omega$, which
	confirms the conjecture~\eqref{scaling}.  In some particular cases, we
	found that $R$ vanishes even faster with small $\Gamma$ or $\Omega$, so
	that we can conclude that Eq.~\eqref{scaling} is a rather conservative
	estimate.  For particular $R$'s (e.g.\ for $R_1^1$ as discussed above) or
	particular systems, the scaling may even be more favorable.  As an example,
	we present in Appendix~C results for the quadruple quantum dot.
	
	As function of the temperature $k_BT=1/\beta$, the deviations behave even
	more interestingly, because they vanish in both the high-temperature and
	the low-temperature limit [see Fig.~\ref{fig:R}(c)].  For the
	high-temperature limit $\beta\to 0$, this is expected since the
	substitution $\xi\to =-\xi+i\beta$ by and large cures the
	fluctuation theorem violation, while being irrelevant for $\beta=0$.
	Quantitatively we find the scaling $R\propto \beta^3$ or even higher
	powers.
	For the experimentally rather relevant low-temperature limit
	$\beta\to\infty$, we find that the deviations turn rather rapidly to zero,
	but do not follow a power law.  Once $k_BT\lesssim \Gamma/10,\Omega/10$,
	all deviations from $R=0$ are already many orders smaller than the
	individual terms of $R$.
	
	\paragraph{Energy fluctuations}
	\begin{figure}[tb]
		\includegraphics{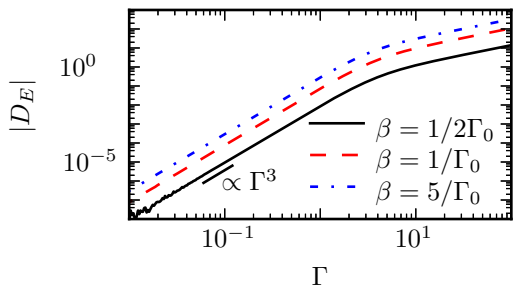}\caption{Fluctuations of the total lead energy manifest in the energy
		diffusion constant $D_E=\lim_{t\to\infty}\langle \Delta
		E^2_\text{leads}\rangle/t$ as function of the dot-lead tunnel rate $\Gamma$
		and various temperatures. The chemical potentials at the upper quantum dot
		are $\mu_1=-\mu_2=3\,\Gamma$, while all other parameters are as in
		Fig.~\ref{fig:R}(a).
		}
		\label{fig:E}
	\end{figure}%
	In the exact treatment, the total energy is conserved while the central
	system can only ingest a finite amount.  Therefore, cumulants of the lead
	energy cannot grow indefinitely, so that the energy current cumulants must
	vanish \cite{Saito2008a}.  For the RWA master equation, they vanish as well
	owing to the $\xi$-independence of the generating function, see discussion
	after Eq.~\eqref{w}.  Beyond RWA this need not be the case, because the
	full Bloch-Redfield equation allows electrons to lose coherence while
	residing on the central system.  Such coherence loss can cause transitions
	between states with different energy, e.g., between bonding and
	anti-bonding states.  Therefore the variance of the total lead energy might
	grow diffusively, as is confirmed by the results shown in Fig.~\ref{fig:E}.
	The scaling with the dot-lead rate is $\propto\Gamma^3$, i.e., equal to
	that of the generic deviations from $R=0$.  For the usual dot-lead models,
	this seems to be a consequence of the approximations underlying the
	Bloch-Redfield equation.
	
	\paragraph{Conclusions}
	By studying exchange fluctuation theorems for quantum transport, we
	have identified a class of master equations for which these theorems hold
	exactly.  Equations of this class are characterized by an equivalence to a
	RWA master equation in a many-body basis for which we proved the validity
	of the fluctuation theorem.  The many-body aspect is rather crucial for
	the direct application to coupled quantum dots given that Coulomb
	interaction represents the largest energy scale in these systems.
	Interestingly, various previous studies \cite{Sanchez2010a,
	BulnesCuetara2011a, Golubev2011a, BulnesCuetara2013a} represent special
	cases of our more generic statements.
	
	Despite that the RWA version of the Bloch-Redfield master equation obeys
	the fluctuation theorem \eqref{FT} exactly and, thus, possesses a desirable
	formal property, it is not necessarily the preferential choice, because
	coherences may be of the same order as the populations so that neglecting
	coherences may lead to even qualitatively wrong predictions
	\cite{Kaiser2006a}.  Going beyond RWA, we quantified the degree of
	fluctuation theorem violation of the full Bloch-Redfield master equation,
	in particular its scaling behavior as function of the coherent and the
	incoherent tunneling.  Most important for the application of the
	Bloch-Redfield master equation to real experiments is the fact that at low
	temperatures, the discrepancies become rather tiny.
	
	Even though our investigation already provides a general proof for the
	consistency of a whole class of master equations with exchange fluctuations
	theorems, two further generalizations seem desirable.  On the one hand, one
	should consider also spin effects, which requires a refined treatment of
	time-reversal symmetry \cite{Lopez2012a}.  On the other hand, one may
	include quantum dissipation for which in the absence of electron
	reservoirs, similar conclusions about the compliance of master equations
	with fluctuation theorems can be drawn \cite{Thingna2012a}, while for the
	combination of transport and dissipation the fluctuation theorem holds at
	least to some extent \cite{Krause2011a}.
	
	We thank R.~S\'anchez for helpful discussions.
	This work was supported by the Spanish Ministry of Economy and
	Competitiveness via grant No.\ MAT2011-24331 and a FPU scholarship (R.H.).

	\appendix
	\section{Iterative computation of transport coefficients}
	
	In order to compute transport coefficients, we adapt the method
	developed in Ref.~\cite{Flindt2008a} for the computation of current
	cumulants.  It is based on two facts: First, for a master equation, the
	zero-frequency current cumulant generating function is given by the
	eigenvalue of the generalized Liouvillian $\mathcal{L}_\chi$ with the
	smallest real value, where $\chi$ is the counting variable
	\cite{Bagrets2003a}.  Second, the cumulants are the Taylor coefficients
	appearing in the expansion of the generating function $Z(\chi)$.  Since
	Rayleigh-Schr\"odinger perturbation theory \cite{Sakurai} provides a series
	expansion of eigenvalues, it can be used to iteratively compute cumulants.
	
	In our case, we have to generalize this method in two respects.  On the one
	hand, we like to compute also energy exchange cumulants which requires
	additional counting variables $\xi_\alpha$ for each lead $\alpha$.  On the
	other hand, we are interested in the transport coefficients, i.e., in a
	series expansion in the chemical potentials of the leads, $\mu_\alpha$,
	around their equilibrium value $\mu_0$ which we set to zero for ease of
	notation.  While the formal aspects of the iteration scheme are the same as
	in its original version, the required series expansion of the Liouvillian
	in the variables $\bm\chi$, $\bm\xi$, and $\bm\mu$ is no longer that of a
	simple exponential.
	
	Following the idea of Ref.~\cite{Flindt2008a}, we start by writing the
	generalized Liouvillian \eqref{L} as series in all these variables,
	\begin{equation}
		\label{app:Lseries}
		\mathcal L_{\bm\chi,\bm\xi,\bm\mu}
		= \mathcal{L} +
		\sum_\alpha \sum_{k,k',k''=0}^\infty \frac{i^{k+k'}}{k!\, k'!\, k''!}
		{\mathcal W}_{k,{k'},{k''}}^\alpha
		\chi_\alpha^k \xi_\alpha^{k'} \mu_\alpha^{k''} ,
	\end{equation}
	with the Taylor coefficients $\mathcal{W}_{0,0,0}^\alpha=0$ and
	$\mathcal{W}_{k,k',k''}^\alpha = \mathcal{W}_{k,k',k''}^{\alpha,\textrm{in}}
	+ \mathcal{W}_{k,k',k''}^{\alpha,\textrm{out}}$, while for $k''>0$
	\begin{align}
{\mathcal W}_{k,{k'},{k''}}^{\alpha,\textrm{out}}
	={}& \partial_{i\chi_\alpha}^k\partial_{i\xi_\alpha}^{k'}\partial_{\mu_\alpha}^{k''}
	\mathcal L_{\bm\chi,\bm\xi,\bm\mu}^\textrm{out}\big|_{\bm\chi,\bm\xi,\bm\mu=\bm0}\nonumber\\
	={}& \int d\tau\; \partial_{i\xi_\alpha}^{k'} \partial_{\mu_\alpha}^{k''}
	F_\alpha^{>}(\tau-\xi_\alpha)\big|_{\bm\xi,\bm\mu=\bm0}
\nonumber\\
		& \times
				\begin{cases}
			J_{n_\alpha}^\textrm{out}(\tau)-D_{n_\alpha}^\textrm{out}(\tau) & \text{for $k=k'=0$} \\
			J_{n_\alpha}^\textrm{out}(\tau) & \text{else}
				\end{cases}
	\end{align}
	with the superoperators
	\begin{align}
	J_{n_\alpha}^\textrm{out}(\tau)\rho & =
		\frac{1}{2}
		\big\{\tilde c_{n_\alpha}(-\tau)\rho c_{n_\alpha}^{\dagger}
		+ c_{n_\alpha}\rho\tilde c_{n_\alpha}^{\dagger}(\tau)\big\}, \\
	D_{n_\alpha}^\textrm{out}(\tau)\rho & =
		\frac{1}{2}
		\big\{c_{n_\alpha}^{\dagger}\tilde c_{n_\alpha}(-\tau)\rho
		+     \rho\tilde c_{n_\alpha}^{\dagger}(\tau)c_{n_\alpha}\big\}.
	\end{align}
	The latter appear in the integrals that provide the jump operators and the
	dissipator, respectively, of the Liouvillian.
	$\mathcal{W}_{k,k',k''}^{\alpha,\textrm{in}}$ follows from the substitution
	$\{c_n,F^{>}(t)\} \to \{c_n^\dagger,F^{<}(-t)\}$ and multiplication by a
	factor $(-1)^{k}$.  Notice that no cross terms between different leads
	emerge.  Furthermore, we separate the Liouvillian into $\mathcal L\rho
	=-i\com{H_S}{\rho}+\mathcal L^{\textrm{in}}\rho+\mathcal
	L^{\textrm{out}}\rho$ where $H_S$ refers to the system Hamiltonian, while
	$\mathcal L^{\textrm{in}}$ and $\mathcal L^{\textrm{out}}$ are the terms in
	the master equation \eqref{ME} that contain the lead correlation functions
	$F^{<}$ and $F^{>}$, respectively.
	
	The derivatives with respect to the heat counting variables $\xi_\alpha$
	and the chemical potentials $\mu_\alpha$ act upon the lead correlation
	functions as
	\begin{align}
& \frac{\partial^{k'}}{\partial(i\xi_\alpha)^{k'}}
  \frac{\partial^{k''}}{\partial\mu_\alpha^{k''}}
  F_\alpha^{>}(\tau-\xi_\alpha)\Big|_{\bm\xi=\bm\mu=\bm{0}}\nonumber\\
&=\frac{\Gamma_\alpha}{2\pi} \int d\tau\;
  e^{-iE\tau}E^{k'}\frac{\partial^{k''}}{\partial\mu_\alpha^{k''}}
  [1-f(E-\mu_\alpha)]\Big|_{\bm\mu=\bm0},
	\end{align}
	where we have restricted ourselves to the wideband
	limit, $F_\alpha^{>}(\epsilon) = \Gamma_\alpha[1-f(\epsilon-\mu_\alpha)]$ with
	the Fermi function $f(E-\mu)=\{\exp[\beta(E-\mu)] +1\}^{-1}$.  Its
	derivatives at equilibrium chemical potential can be expressed as series,
	\begin{equation}
		\begin{split}
			\frac{\partial^{k''}}{\partial\mu^{k''}}f(E-\mu)\Big|_{\mu=0}
			={}& (-\beta)^{k''}\sum_{m=0}^{k''} (-1)^m m! \;S_{k'',m}  \\
			& \times [1-f(E)]^m [f(E)]\label{app:FermiSeries}
		\end{split}
	\end{equation}
	with $S_{k'',m}$ the Stirling numbers of the second kind
	\cite{Gradshteyn1994a}. To derive this formula, we start with the
	expression $\partial_x^n(e^x+1)^{-1}$ and employ Fa\`{a} di Bruno's Formula
	\cite{JohnsonAMMonthly2002a} for the derivative of nested functions.  Exploiting a relation
	between Stirling numbers and the partial Bell polynomials,
	$B_{n,k}(e^x,\ldots,e^x)=e^{kx}S_{n,k}$ yields
	\begin{align}
\frac{\partial^n}{\partial x^n} \frac{1}{e^x+1}
&= \sum_{k=0}^{n} (-1)^k k! \;S_{n,k}\frac{e^{kx}}{(e^x+1)^{k+1}},
	\end{align}
	by which we immediately obtain the $n$th derivative of the Fermi function
	with respect to the chemical potential and, hence, the Taylor
	series~\eqref{app:FermiSeries}.
	
	Finally, we end up with an eigenvalue problem that is equivalent to the
	one of Ref.~\cite{Flindt2008a} but with the additional perturbations
	$\bm\xi$ and $\bm\mu$.  Despite that the coefficients now look more
	involved, the iteration scheme derived there can be applied
	straightforwardly.
	
	\section{Johnson-Nyquist relation}
	\begin{figure}[tb]
		\includegraphics{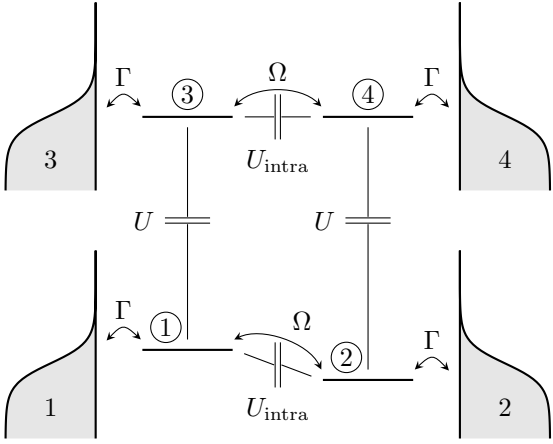}\caption{Quadruple quantum dot in contact with four leads
		$\alpha=1,\ldots,4$.  The system can be considered as two coupled transport
		channels, each formed by a double quantum dot and interacting capacitively
		with the other.
		\label{fig:sketch_QQD}
		}
	\end{figure}%
	\begin{figure*}[tb]
		\includegraphics{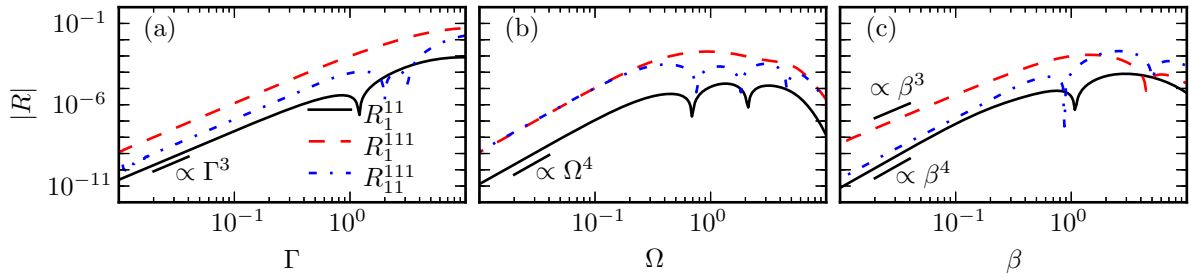}\caption{Deviation of the exchange fluctuation theorem for the quadruple
		quantum dot as function of (a) the dot-lead coupling $\Gamma$, (b) the
		inter-dot tunneling $\Omega$, and (c) the inverse temperature
		$\beta=1/k_BT$ for the default parameters
		$\Gamma=0.5\,\Omega=k_BT=10\epsilon_1=-10\epsilon_2$ and $\mu_\alpha=0$.
		The selected generalized Casimir-Onsager relations are those of
		Fig.~\ref{fig:R}, i.e., $R_{1}^{11}=0$ (solid line), $R_{1}^{111}=0$
		(dashed), and $R_{11}^{111}=0$ (dash-dotted).
		}
		\label{fig:R_QQD}
	\end{figure*}%
	Even though the Bloch-Redfield master equation beyond RWA does not fulfill
	the fluctuation theorem exactly, the resulting conductance
	$G_{\alpha,\alpha} = -\partial I_\alpha/\partial\mu_\alpha|_{\bm\mu=\bm0}$
	and the zero-frequency noise $S_{\alpha\alpha} = \partial^2
	Z/\partial\chi_\alpha^2|_{\bm\chi=\bm\xi=\bm0}$ at equilibrium
	nevertheless obey the Johnson-Nyquist relation $2G_{\alpha,\alpha}=\beta
	S_{\alpha\alpha}$.  For a proof, we perform the iteration
	described above up to second order which yields the expressions
	\begin{align}
\label{app:S}
	S_{\alpha\alpha} &= \ev{{\mathcal W}_{2,0,0}^\alpha}
	+2\ev{{\mathcal W}_{1,0,0}^{\alpha}\mathcal R{\mathcal W}_{1,0,0}^\alpha} ,
\\
\label{app:G}
	G_{\alpha,\alpha} &= -\ev{{\mathcal W}_{1,0,1}^\alpha}
	-\ev{{\mathcal W}_{1,0,0}^\alpha\mathcal R{\mathcal W}_{0,0,1}^\alpha}
	-\ev{{\mathcal W}_{0,0,1}^{\alpha}\mathcal R{\mathcal W}_{1,0,0}^{\alpha}},
	\end{align}
	where the angular brackets denote the expectation value with respect to the
	grand canonical density operator of the central system, $\rho_{\textrm{eq}}
	\propto e^{-\beta(H_S -\mu_0 N)}$.  Notice that $\rho_\text{eq}$ is the
	equilibrium solution of the Bloch-Redfield master equation \eqref{ME},
	i.e., $\mathcal{L}\rho_\text{eq} = 0$ if
	all lead chemical potential are equal, $\mu_\alpha=\mu_0$ (see
	remark at the end of this section).  Here, $\mathcal R =
	-(\mathcal{QLQ})^{-1}$ denotes the pseudo resolvent of the Liouvillian at
	zero frequency (i.e.\ $-\mathcal{R}$ is the pseudo inverse) with
	$\mathcal{Q} = (\openone-\rho_{\textrm{eq}}\tr)$ the projector to the
	Liouville subspace orthogonal to the equilibrium density operator.
	
	We proceed by showing that in Eq.~\eqref{app:G}, the first two terms obey
	the relations $2\ev{{\mathcal W}_{1,0,1}^{\alpha} }= -\beta \ev{ {\mathcal
	W}_{2,0,0}^\alpha}$ and ${\mathcal W}_{0,0,1}^{\alpha}\rho_{\textrm{eq}} =
	-\beta {\mathcal W}_{1,0,0}^{\alpha}\rho_{\textrm{eq}}$, respectively,
	while the last term vanishes, $\tr{\mathcal W}_{0,0,1}^{\alpha}=0$.
	The latter relation follows from the fact that the trace condition of
	the Liouvillian is independent of the lead chemical potential, so that the
	corresponding Taylor expansion vanishes to all orders.
	
	The proof for the other two relations is more involved. It is on
	the Kubo-Martin-Schwinger relation for the lead correlation functions
	\cite{Kubo1957a, Martin1959a},
	\begin{align}
	F_\alpha^{>}(t) = e^{-\beta\mu_\alpha}\; F_\alpha^{<}(t+i\beta). \label{app:DetBalF}
	\end{align}
	and a related detailed balance relation for the interaction picture
	operators,
	\begin{align}
\tilde c_{n_\alpha}(t) e^{-\beta(H_S-\mu_0 N)}
&=e^{\beta\mu_0} e^{-\beta(H_S-\mu_0 N)}\tilde c_{n_\alpha}(t-i\beta) .
	\end{align}
	The latter holds for fermionic annihilation operators in the interaction
	picture with respect to $H_S$, i.e., for any $\tilde c_{n_\alpha}(t) =
	e^{iH_St}c_{n_\alpha} e^{-iH_St}$ of the system, owing to the commutator
	$[N,c_{n_\alpha}] = -c_{n_\alpha}$.  From this relation follow
	detailed balance relations for the jump operators,
	\begin{align}
	D_{n_\alpha}^\textrm{in}(t) \rho_{\textrm{eq}}
	&= e^{-\beta\mu_0}J_{n_\alpha}^\textrm{out}(-t-i\beta) \rho_{\textrm{eq}}, \label{app:DetBalJOut} \\
	D_{n_\alpha}^\textrm{out}(t) \rho_{\textrm{eq}}
	&= e^{\beta\mu_0}J_{n_\alpha}^\textrm{in}(-t-i\beta) \rho_{\textrm{eq}}, \label{app:DetBalJIn}
	\end{align}
	which we use to transform the superoperators appearing in $S_{\alpha\alpha}$.
	
	We start with the tunnel-out contribution of the first term of Eq.~\eqref{app:G},
	\begin{align}
\tr\mathcal W_{1,0,1}^{\alpha,\textrm{out}}\rho_{\textrm{eq}}
={}&\tr\int d\tau\;
	\frac{\partial}{\partial\mu_\alpha}F_\alpha^{>}(\tau)J_{n_\alpha}^\textrm{out}(\tau)
	\Big|_{\mu_\alpha=\mu_0}\rho_{\textrm{eq}} ,
	\end{align}
	insert Eqs.~\eqref{app:DetBalF}, \eqref{app:DetBalJOut}, \eqref{app:DetBalJIn} and substitute the
	integration variable $\tau \to -\tau-i\beta$.  Again we use that $\mathcal
	W_{0,0,1}^{\alpha,\text{out}}$ is trace free and obtain
	\begin{align}
	\ev{\mathcal W_{1,0,1}^{\alpha,\textrm{out}}}
	&= -\beta\ev{\mathcal W_{2,0,0}^{\alpha,\textrm{in}}}-\ev{\mathcal W_{1,0,1}^{\alpha,\textrm{in}}}.
	\end{align}
	This relation together with the corresponding expression for the tunnel-in
	term, $\ev{\mathcal W_{1,0,1}^{\alpha,\textrm{in}}} = -\beta\ev{\mathcal
	W_{2,0,0}^{\alpha,\textrm{out}}}-\ev{\mathcal
	W_{1,0,1}^{\alpha,\textrm{out}}}$, yields $\ev{{\mathcal
	W}_{1,0,1}^{\alpha} }= -(\beta/2) \ev{ {\mathcal W}_{2,0,0}^\alpha}$, which
	links the first term of Eq.~\eqref{app:S} to that of Eq.~\eqref{app:G}.
	
	Following the same path for the second term, we find
	\begin{align}
\mathcal W_{0,0,1}^{\alpha,\textrm{out}}\rho_{\text{eq}}
={}&\int \!\! d\tau\;
	\partial_{\mu_\alpha}
	F_\alpha^{>}(\tau)[J_{n_\alpha}^\textrm{out}(\tau)
	-D_{n_\alpha}^\textrm{out}(\tau)\Big|_{\mu_\alpha=\mu_0}\!\!\!\rho_{\textrm{eq}}
\nonumber\\
={}& \big[\mathcal W_{1,0,1}^{\alpha} - \beta \mathcal
     W_{1,0,0}^{\alpha,\textrm{in}}\big]\rho_{\textrm{eq}} ,
\label{app:F}
	\end{align}
	as well as $\mathcal W_{0,0,1}^{\alpha,\textrm{in}}\rho_{\textrm{eq}} =
	(-\mathcal W_{1,0,1}^{\alpha} - \beta \mathcal
	W_{1,0,0}^{\alpha,\textrm{out}})\rho_{\textrm{eq}}$.  Thus, also the
	second terms in Eqs.~\eqref{app:S} and \eqref{app:G} differ only by a factor
	$\beta/2$, which completes our proof that the conductivity and the
	zero-frequency noise computed with the full Bloch-Redfield master equation
	\eqref{ME} obey the Johnson-Nyquist relation $S_{\alpha\alpha} = 2 k_BT
	G_{\alpha,\alpha}$.
	
	Finally, let us remark that Eqs.~\eqref{app:DetBalF}, \eqref{app:DetBalJOut}  and
	\eqref{app:DetBalJIn} can be used to demonstrate that the grand canonical
	state of the central system, $\rho_\text{eq} \propto
	\exp[-\beta(H_S-\mu_0N)]$, represents the equilibrium solution of the
	Bloch-Redfield master equation \eqref{ME} both within RWA and beyond.
	Thus, $\mathcal{L}\rho_\text{eq}=0$ and
	$\mathcal{L}_\text{RWA}\rho_\text{eq}=0$ provided that no bias voltages are
	applied so that all lead chemical potentials are equal.
	As further consequence, for both master equations the current vanishes at
	equilibrium as expected.
	
	\section{Numerical results for a quadruple quantum dot}
	
	\begin{figure}[tb]
		\includegraphics{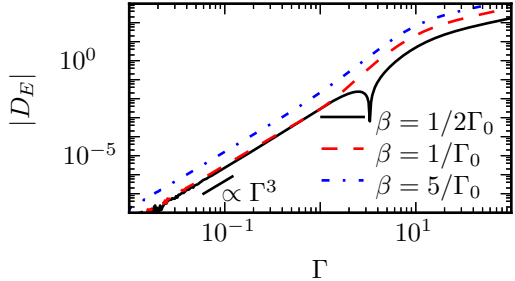}\caption{\label{fig:E_QQD}
		Diffusion constant $D_E=\lim_{t\to\infty}\langle \Delta E^2_\text{leads}\rangle/t$
		of the total lead energy for the setup sketched in Fig.~\ref{fig:sketch_QQD}
		as function of the dot-lead tunnel rate $\Gamma\equiv\Gamma_\alpha$
		and various temperatures.  The chemical potential of leads 1 and 2 read
		$\mu_1=-\mu_2=3\,\Gamma$, while all other parameters are as in
		Fig.~\ref{fig:R_QQD}(a).
		}
	\end{figure}%
	
	As a special system for which the deviations from the fluctuation theorem
	scale even more favorable as the behavior given by Eq.~\eqref{R}, we
	present numerical results for a quadruple quantum dot coupled to four
	leads, as is sketched in Fig.~\ref{fig:sketch_QQD}.  The deviation from
	$R=0$ as function of the dot-lead coupling $\Gamma$ and the inverse
	temperature $\beta=1/k_BT$ [Figs.~\ref{fig:R_QQD}(a) and
	\ref{fig:R_QQD}(c)] is the generic one, i.e., $R\propto \Gamma^3$, while
	$R$ vanishes in the high-temperature limit $\propto\beta^3$ or faster.  In
	the low-temperature limit $\beta\to\infty$, the deviations decay rapidly
	without following any power law.  Also for the lead energy variance behaves
	generically, as can be appreciated in Fig.~\ref{fig:E_QQD}.
	The main difference to the generic behavior is found as function of the
	coherent inter-dot tunneling $\Omega$: We observe a decay
	$R\propto\Omega^4$, i.e., faster than the generic $\propto\Omega^2$
	discussed in the main text.

\end{document}